\documentclass[conference]{IEEEtran}
\usepackage{cite}
\usepackage{amsmath,amssymb,amsfonts}
\usepackage{algorithmic}
\usepackage{graphicx}
\usepackage{textcomp}
\usepackage{xcolor}

\usepackage{hyperref}
\usepackage{float}
\usepackage{booktabs} 
\usepackage{array}    
\usepackage{tabularx} 
\usepackage{geometry} 
\usepackage{listings}  

\def\BibTeX{{\rm B\kern-.05em{\sc i\kern-.025em b}\kern-.08em
    T\kern-.1667em\lower.7ex\hbox{E}\kern-.125emX}}
\begin{document}

\title{Optimizing Recommendations using Fine-Tuned LLMs
\thanks{This paper was accepted and presented at IEEE CAI 2025.}
}
\author{
\IEEEauthorblockN{Prabhdeep Cheema}
\IEEEauthorblockA{\textit{Whiting School of Engineering EP Program} \\
\textit{John Hopkins University}\\
Laurel, MD USA \\
pcheema2@jhu.edu}
\and
\IEEEauthorblockN{Erhan Guven}
\IEEEauthorblockA{\textit{Whiting School of Engineering EP Program} \\
\textit{John Hopkins University}\\
Laurel, MD USA \\
eguven2@jhu.edu}
}

\maketitle
\begin{center}
\textit{This is a preprint of a paper accepted and presented at IEEE CAI 2025.}
\end{center}

\begin{abstract}
As digital media platforms strive to meet evolving user expectations, delivering highly personalized and intuitive movies and media recommendations has become essential for attracting and retaining audiences. Traditional systems often rely on keyword-based search and recommendation techniques, which limit users to specific keywords and a combination of keywords. This paper proposes an approach that generates synthetic datasets by modeling real-world user interactions, creating complex chat-style data reflective of diverse preferences. This allows users to express more information with complex preferences, such as mood, plot details, and thematic elements, in addition to conventional criteria like genre, title, and actor-based searches. In today's search space, users cannot write queries like ``Looking for a fantasy movie featuring dire wolves, ideally set in a harsh frozen world with themes of loyalty and survival.''

Building on these contributions, we evaluate synthetic datasets for diversity and effectiveness in training and benchmarking models, particularly in areas often absent from traditional datasets. This approach enhances personalization and accuracy by enabling expressive and natural user queries. It establishes a foundation for the next generation of conversational AI-driven search and recommendation systems in digital entertainment.
\end{abstract}


\section{Introduction}
As digital media platforms grow rapidly, media companies find that effective search and recommendation systems are essential for attracting and retaining users. With the expansion of streaming libraries and several streaming companies opening their own production houses, users increasingly desire content that aligns with their tastes. Still, traditional recommendation systems often fail to cater to these individualized needs. Current search and recommendation models generally depend on keyword-based approaches, which limit users to set genres, actors, or titles. These models have a narrow focus and struggle to capture the complex and subtle preferences of users, such as mood, plot details, or distinct thematic elements, along with the interplay of items such as genres, actors, or titles. For example, users looking for a ``fantasy movie featuring dire wolves, ideally set in a harsh frozen world with themes of loyalty and survival'' may struggle to communicate their interests using standard keyword searches, resulting in less satisfying and relevant recommendations.

This study evaluates methods to overcome these challenges by generating synthetic datasets replicating more sophisticated user interactions where none exist because the major streaming companies do not yet use AI Assistants. By creating chat-style data rooted in genuine user intents, our model facilitates complex queries, enabling users to describe their content preferences in detail while combining standard search criteria with expressive details. The dataset generation utilizes advanced LLMs, like the Llama3.1 model featuring 405 billion parameters and knowledge graphs, which allows for creating highly realistic and varied queries that emulate natural user behavior. This is especially important because there is a lack of such datasets. Previous work in \cite{lu2023august} shows techniques for generating such datasets leveraging knowledge graphs and user interactions, while \cite{10020479} utilized a Generative Adversarial Network (GAN).

We utilize models with fewer than 70 billion parameters to strike an optimal balance between computational efficiency and performance. Through fine-tuning these models on targeted tasks, we enhance their capability to manage intricate search and recommendation queries effectively. To facilitate this, we employ Low-Rank Adaptation (LoRA) \cite{hu2021lora} and Quantized LoRA (QLoRA) \cite{dettmers2023qlora} techniques, which enable the adaptation of high-parameter models while minimizing computational overhead. These methods provide a cost-effective and robust solution for integrating new information into the model, significantly enhancing its ability to process sophisticated and expressive user queries.

\subsection {Objective and Contribution}
In this study, we develop and evaluate a framework that enhances media recommendation systems using synthetic datasets and fine-tuned large language models, enabling personalized and expressive conversational search capabilities. This includes leveraging structured data enrichment via knowledge graphs, optimizing LLMs with cost-effective techniques like LoRA and QLoRA, and demonstrating the impact of these methods on improving accuracy, diversity, and contextual relevance in recommendations. Key contributions of this work include:
(1) a novel framework for generating synthetic datasets tailored to conversational search and recommendations; this allows the generation of conversational data for specific information retrieval tasks when there is none or little actual data.
(2) leveraging existing knowledge graphs to enrich datasets with structured, domain-specific attributes, improving contextual accuracy in recommendations.
(3) fine-tuning Large Language Models using cost-effective LoRA and QLoRA techniques.
(4) benchmarking the diversity and impact of these datasets in search and recommendation tasks.
(5) utilizing smaller, faster models to match the milli-second response times needed for search and recommendation tasks.

\section{Related Work}
Exploring LLMs as the basis for advanced search and recommendation systems has sparked considerable interest in recent studies. Conventional recommendation systems typically emphasize static, content-based interactions, often failing to deliver interactive, user-focused suggestions. Incorporating LLMs in this area aims to fill this void by emphasizing the creation of conversational, adjustable recommendations that adapt to changing user preferences. This change shows how LLMs can improve recommendation systems. Instead of using fixed keywords, these models can understand and respond to complex user input. A significant work in this field is \cite{lu-etal-2024-aligning}. The authors introduce a novel strategy that aligns LLMs to offer controllable recommendations. The research outlines supervised learning with annotated recommendation tasks and reinforcement learning-based alignment to tackle alignment issues. Through comprehensive experiments, they illustrate that this method enhances the relevance of LLM responses and compliance with user-directed guidance, tackling prevalent challenges such as insufficient personalization and errors in response formatting.

A recent paper \cite{Xu2024Magpie} presents experiments demonstrating that models fine-tuned with synthesized data perform comparably to those fine-tuned with extensive supervised datasets. The authors name their strategy MAGPIE. This finding supports our strategy of potentially using synthetic data to enhance knowledge-enriched LLMs' retrieval and recommendation performance. The MAGPIE framework was published recently, around the same time we concluded our work on synthetic data techniques. The fact that similar strategies are being validated by contemporary research in the field is encouraging. However, MAGPIE differs from our method in key aspects. Their approach generates general-purpose synthetic data, while our method focuses on targeted data generation grounded in specific ontologies and knowledge graphs. Additionally, we leverage larger models, such as Meta’s 405-billion-parameter LLM, whereas MAGPIE primarily utilizes Meta’s 8B and 70B models. Another key difference lies in the optimization techniques employed. MAGPIE uses Supervised Fine-Tuning (SFT) as well as Direct Preference Optimization (DPO) and a Reward Model (RM) to refine alignment and improve the quality of synthetic data. In contrast, our approach relies solely on SFT. Their integration of DPO and RM has shown promising results, suggesting potential avenues for further improving synthetic data generation.

\section{Method}

\subsection{Datasets}

We start with the movie dataset \cite{asaniczka_themoviedb_org_2024}, the most comprehensive with a fine-grained set of genres. We also use \cite{AmazonPrimeMoviesAndTVShows}, \cite{NetflixShows}, and \cite{AmazonReviewsDataset} to extract titles, genres, cast, directors, themes and plots. All three data sets are also used to create the synthetic prompt. This variation between sources enhances the diversity of user inputs to measure the efficacy of the fine-tuned task specialty. It also helps emulate a more realistic behavior usually shown by online users.

We employ two primary strategies to add diversity and variety to our synthetic data: grammar extension using LLMs and rotating seed data in our synthetic prompt templates. The first strategy involves enhancing generated texts' grammatical structure and complexity through advanced language model capabilities, ensuring a wide range of linguistic variations. The second strategy, rotating seed data, involves systematically altering the initial data inputs for generating prompts. This approach helps simulate a broader spectrum of user queries, enriching the dataset with varied scenarios and contexts. These methods collectively contribute to a robust dataset that supports comprehensive model training and evaluation. We generated a dataset consisting of 20,000 diverse prompts using these techniques.

We utilized Llama 3.1 405B to generate the synthetic data, which is recognized for its quality based on the latest benchmarks from the \href{https://huggingface.co/spaces/lmarena-ai/chatbot-arena-leaderboard}{Chatbot Arena Leaderboard on Hugging Face} and features a less restrictive license. While Llama 3.1 405B is known for its quality because of its massive size with 405 billion parameters, it is one of the slowest and most expensive models to host. Meta also used larger models to create the smaller models using knowledge distillation and pruning \cite{meta_llama3.2_2024}\cite{10.1007/s11263-021-01453-z}\cite{wang-etal-2020-structured} as shown in the Fig.~\ref{fig:model}:

\begin{figure}[b]
\centering
\includegraphics[width=0.5\textwidth]{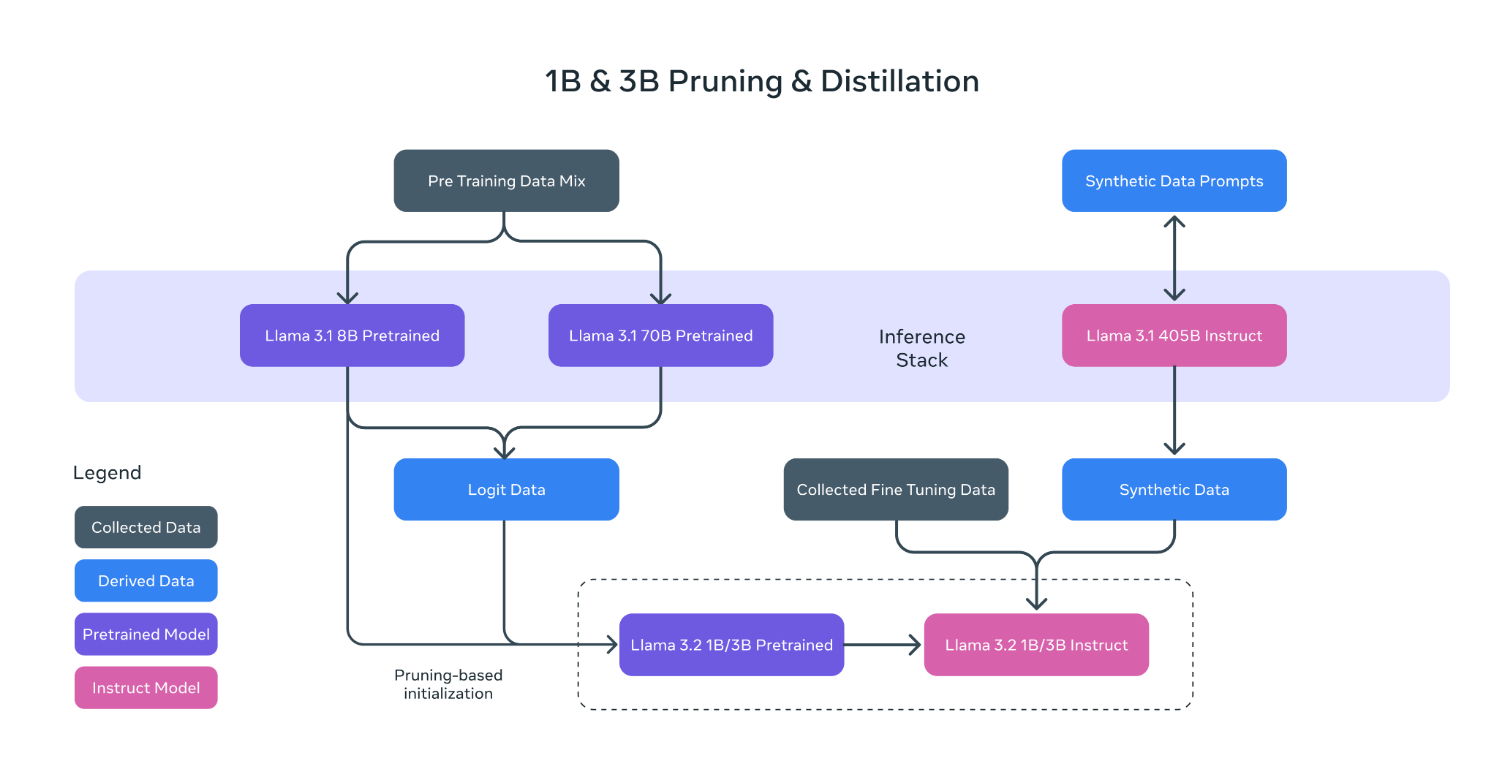}
\caption{Lightweight models using knowledge distillation and pruning
 (\cite{meta_llama3.2_2024})}
\label{fig:model}
\end{figure}

With the present hardware limitations, smaller models such as 3B, 7B, and 8B are better suited for search and recommendations because of their faster response times. These are essential for applications where responses under 500ms are highly desirable. We also tried OpenAI models for Synthetic Data Generation using GPT-4 and GPT-4o; however, these models crashed while generating complicated large synthetic datasets.

\begin{figure}[t]
\centering
\includegraphics[width=0.4\textwidth]{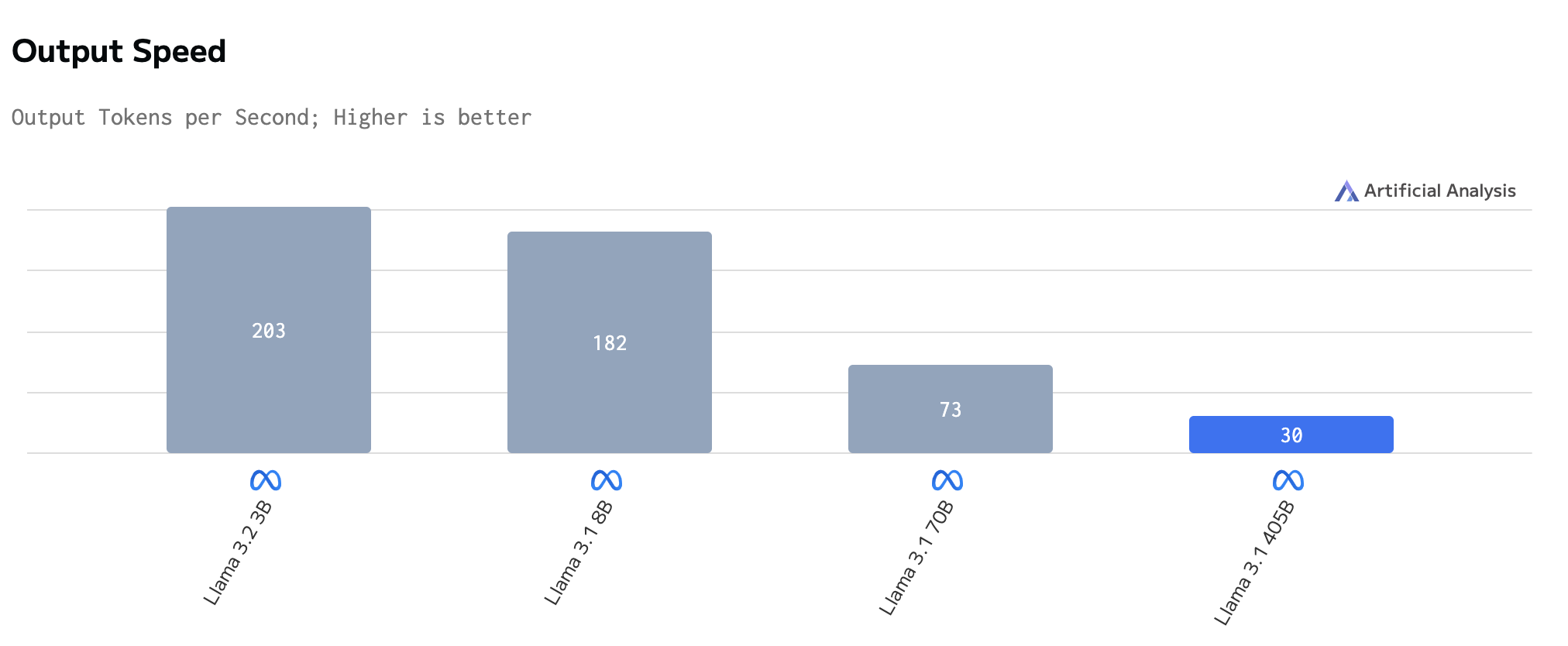}
\caption{Speed
Measured by Output Speed (tokens per second) (\cite{ArtificialAnalysis2024})}
\label{fig:example1}
\end{figure}

\begin{figure}[t]
\centering
\includegraphics[width=0.4\textwidth]{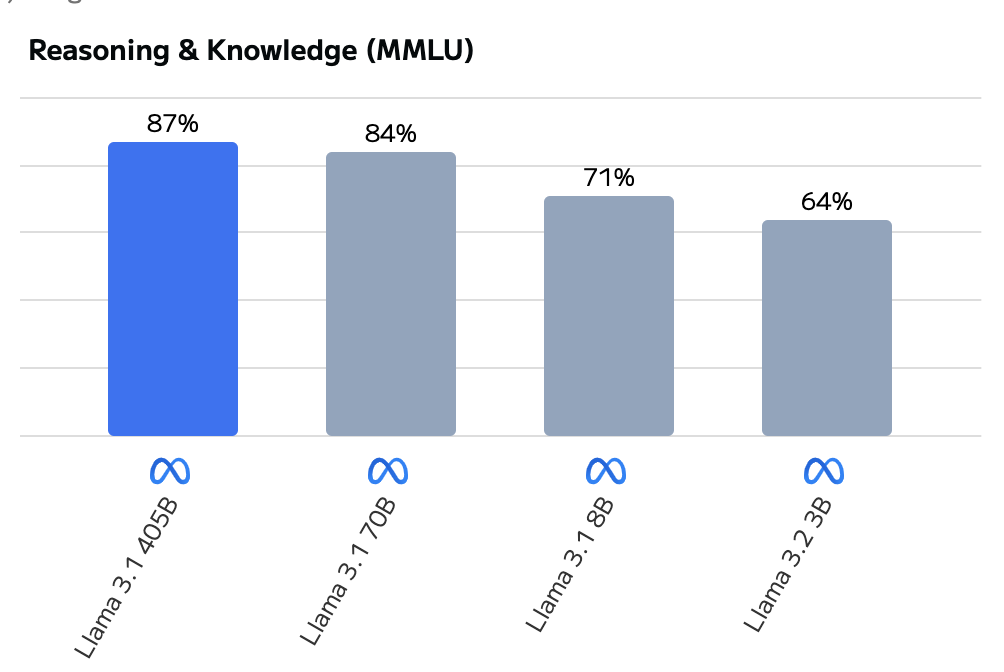}
\caption{Quality based on Multi-task Language Understanding (MMLU) (\cite{ArtificialAnalysis2024})}
\label{fig:example2}
\end{figure}

Following are a few examples of synthetic data grammar:

\begin{enumerate}
    \item ``I loved [movie title]. Any recommendations along those lines?''
    \item ``Can you recommend a movie about [plot theme, e.g., time travel, friendship]?''
    \item ``Are there any good [genre] films set in [specific setting, e.g., a post-apocalyptic world]?''
\end{enumerate}

\subsection{Fine-tuning}
We will use a variant of LoRA \cite{hu2021lora}, one of the most popular and effective methods for efficiently training customized LLMs. This variant is called QLoRA \cite{dettmers2023qlora}. LoRA freezes the weights of pre-trained models and introduces trainable rank decomposition matrices into each layer of the Transformer architecture. This approach significantly reduces the number of trainable parameters required for downstream tasks. Compared to fine-tuning GPT-3 with 175 billion parameters using Adam, LoRA reduces the number of trainable parameters by a factor of 10,000 and decreases the GPU memory requirements by three times. Additionally, LoRA achieves model performance comparable to, or even better than, fine-tuning on models such as RoBERTa, DeBERTa, GPT-2, and GPT-3. This is accomplished while maintaining fewer trainable parameters, a higher training throughput, and, unlike adapters, with no increased inference latency.

\begin{figure}[t]
\centering
\includegraphics[width=0.2\textwidth]{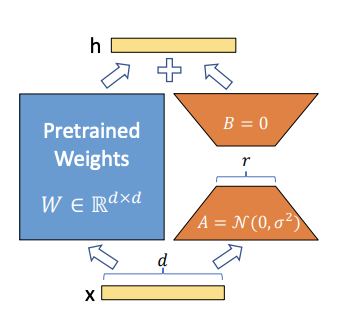}
\caption{LoRA only trains A and B (\cite{hu2021lora})}
\label{fig:example3}
\end{figure}

QLoRA variant is an efficient fine-tuning approach that reduces memory usage enough to fine-tune a 65B parameter model on a single 48GB GPU while preserving full 16-bit fine-tuning task performance. QLoRA backpropagates gradients through a frozen, 4-bit quantized pre-trained language model into Low-Rank Adapters. QLoRA introduces several innovations to save memory without sacrificing performance:

\begin{enumerate}
    \item  4-bit NormalFloat (NF4), a new data type that is information-theoretically optimal for normally distributed weights
    \item double quantization to reduce the average memory footprint by quantizing the quantization constants
    \item paged optimizers to manage memory spikes.
\end{enumerate}

\begin{figure}[b]
\centering
\includegraphics[width=0.5\textwidth]{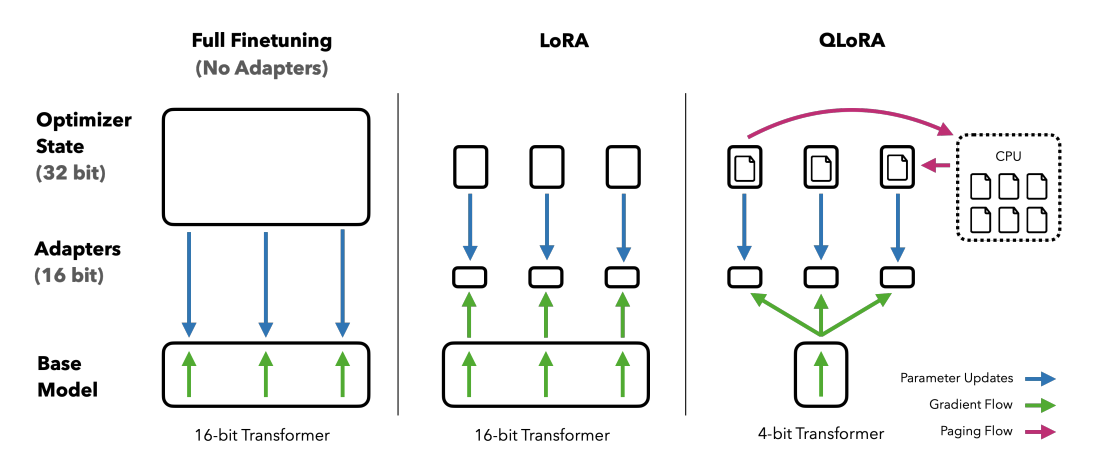}
\caption{QLORA improves over LoRA by quantizing the transformer model to 4-bit precision and using paged optimizers to handle memory spikes (\cite{dettmers2023qlora})}
\label{fig:example4}
\end{figure}

Our fine-tuning tasks will train the LLM to optimize its response for the following specific tasks:

\begin{enumerate}
    \item Extract intent from the user prompts
    \item Extract plots, genres, and themes from the user prompts
\end{enumerate}

As a result of QLoRA we could train Llama 3B and 8B models on single A10 and L4 GPU. We experimented more with Llama3.2 3B due to faster inference and a smaller hardware footprint. Meta mentions in their Llama 3.2 release \cite{meta_llama3.2_2024} that they utilized two techniques — pruning and distillation — on the 1B and 3B models, making them the first highly capable, lightweight Llama models that can efficiently fit on smaller hardware. Llama 3B with FP-32 does not fit on a single A10 or L4 GPU.

\subsection{Training pre-processing and Prompt Templates}
We experimented with Llama and Mistral templates. We concentrated more on Llama3.2 models due to their faster inference. Below is the sample template for the Llama model experiments

Using a proper prompt template and system prompt is crucial to performing the underlying tasks. The training data is formatted using the template in Listing 1. For example, below is the template for the intent tasks.

\begin{lstlisting}[caption=Llama Prompt Template,  basicstyle=\scriptsize]
<|begin_of_text|><|start_header_id|>system<|end_header_id|>\n{instr}<|eot_id|>\n<|start_header_id|>user<|end_header_id|>\n{input}<|eot_id|>\n<|start_header_id|>assistant<|end_header_id|>\n{json_output}<|eot_id|>\n<|end_of_text|>
\end{lstlisting}

\begin{lstlisting}[caption=Intent training data,  basicstyle=\scriptsize]
<|begin_of_text|><|start_header_id|>system<|end_header_id|>\nYou will receive User prompt with questions about Movies. Your task is to classify the intent among items in list ["rec", "non_rec"].Where 'rec' means the prompt is asking for recommendations and 'non_rec' means the prompt is asking about non-recommendations general information about movie related stuff for example:Who directed 'Forrest Gump' and what other films has he made?.Respond in json for example: {"intent": "non_rec"}\n<|eot_id|>\n<|start_header_id|>user<|end_header_id|>\nCan you provide information on the production company behind 'Game of Thrones'?<|eot_id|>\n<|start_header_id|>assistant<|end_header_id|>\n{"intent": "non_rec"}<|eot_id|>\n<|end_of_text|>
\end{lstlisting}

\subsection{LoRA Configuration}

We iterated over all layers using grid search to converge to the optimal set of target layers. The best results were achieved with the following combination:
\begin{lstlisting}[caption=Target layers, basicstyle=\scriptsize]
["q_proj", "v_proj", "o_proj"]
\end{lstlisting}

The table below shows the full list of target layers iterated over:

\begin{table}[t]
\centering
\caption{Description of Various Projection Layers in Transformer Models}
\label{tab:projection_layers}
{\scriptsize 
\begin{tabularx}{\linewidth}{@{}lX@{}}
\toprule
\textbf{Layer} & \textbf{Description} \\ \midrule
\textbf{q\_proj} & Projects input features into the query space of the attention mechanism. \\
\textbf{k\_proj} & Projects input features into the key space for computing relevance scores. \\
\textbf{v\_proj} & Projects input features into the value space, representing content to be aggregated. \\
\textbf{o\_proj} & This is the final projection applied to the weighted sum of the attention mechanism, generating the output of the attention layer. \\
\textbf{up\_proj} & Hierarchical attention mechanisms, where information is projected up.  \\
\textbf{down\_proj} & Hierarchical attention mechanisms, where information is projected. down \\
\textbf{gate\_proj} & Generates a ``gate'' value that controls the flow of information within the attention mechanism. \\ \bottomrule
\end{tabularx}
}
\end{table}

\subsection{Model Training}
The loss graph for both the Entity and Intent extraction fine-tuning tasks shows a consistent decrease in loss, indicating effective learning and model optimization; this decrease slows significantly as training progresses.

\begin{figure}[!t]
\centering
\includegraphics[width=0.4\textwidth]{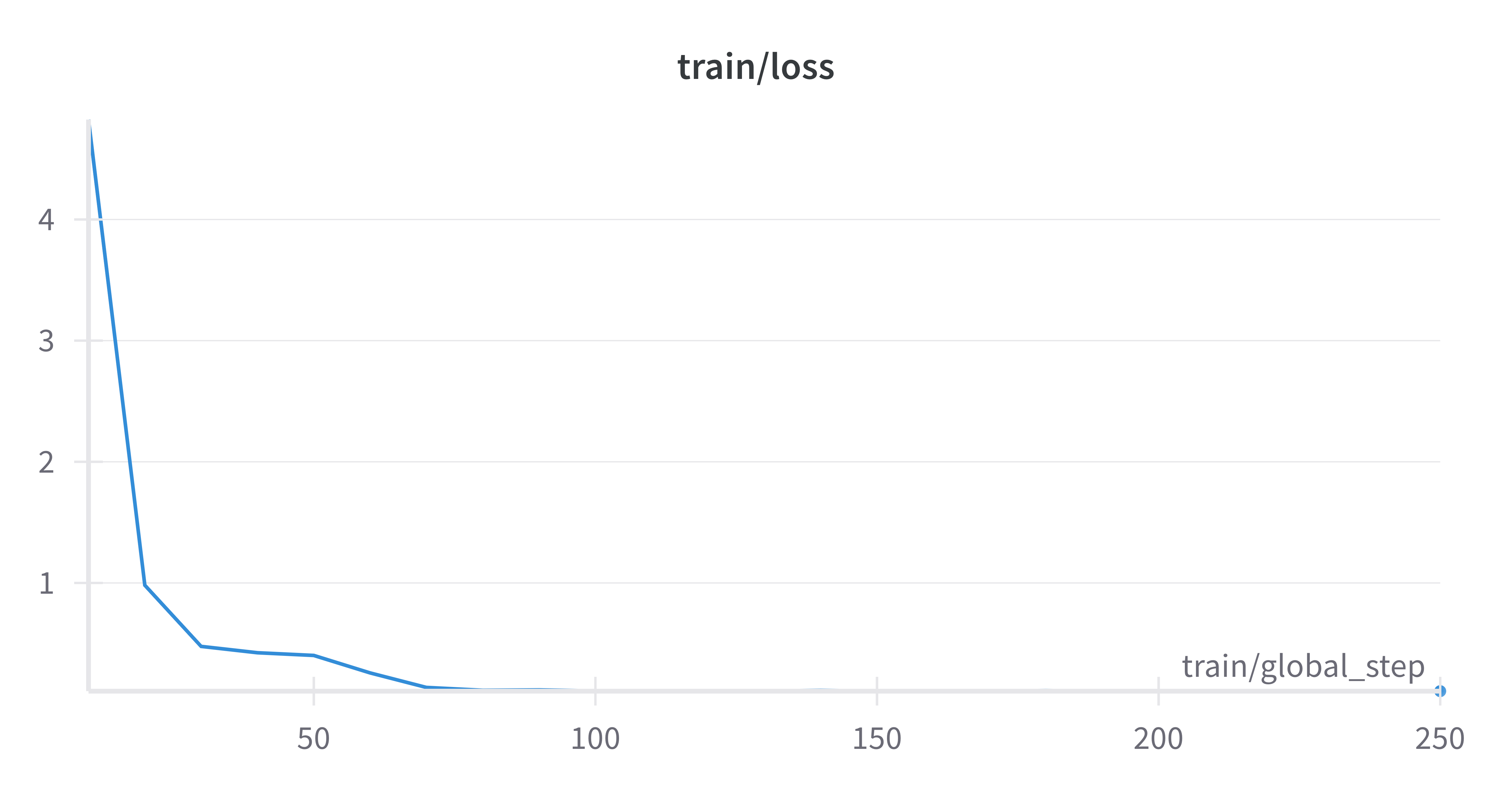}
\caption{Prompt Entity Extraction}
\label{fig:example5}
\end{figure}

\begin{figure}[!t]
\centering
\includegraphics[width=0.4\textwidth]{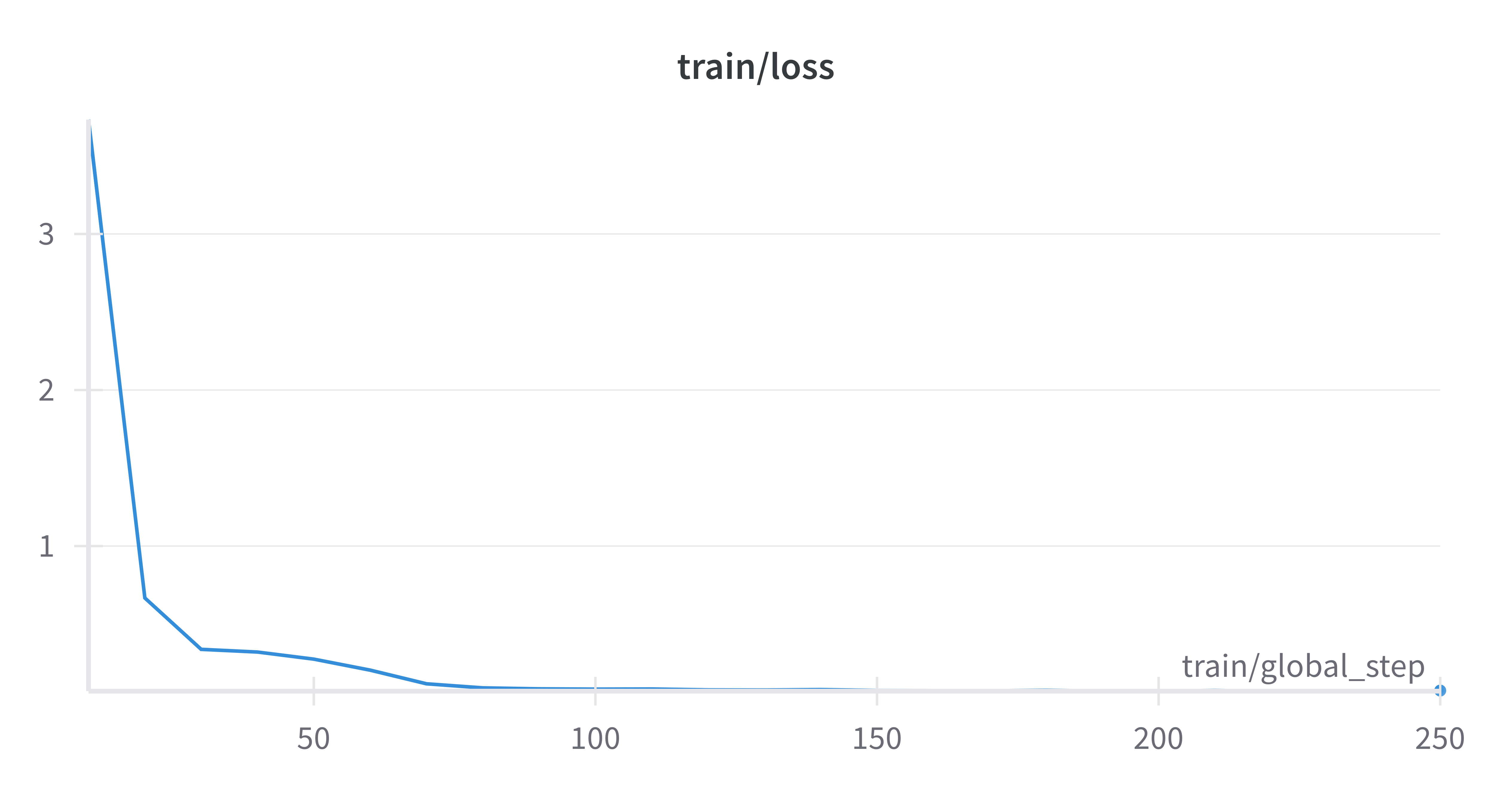}
\caption{Prompt Intent Extraction}
\label{fig:example6}
\end{figure}

\subsection{Evaluation Methodology}

We will use Macro-F1 to evaluate both Entity Extraction and Intent Classification. Macro-F1 is particularly well-suited for these tasks because it provides a balanced performance measure across all classes, regardless of their frequency. By emphasizing performance across all classes, Macro-F1 ensures the evaluation is fair and robust. This makes it a reliable metric for assessing the effectiveness of our model in capturing both entities and intents accurately.

\section{Results}
We compare our fine-tuned models against the vanilla Llama 3.2 3B model (using carefully designed system prompts) and the best-performing Named Entity Recognition (NER) models available on Hugging Face. This comprehensive evaluation allows us to assess how effectively our fine-tuning approach improves performance over baseline models and how it competes with state-of-the-art NER systems. We benchmark our results against widely recognized, highly optimized systems by including Hugging Face's top-performing NER models, ensuring a robust and fair evaluation. This highlights the relative strengths of our approach in extracting entities across various classes.

We evaluate our fine-tuned models against the vanilla Llama 3.2 3B model for intent classification using carefully designed system prompts. These prompts are tailored to elicit the most accurate responses from the baseline models, ensuring a fair and controlled comparison. This configuration enables us to thoroughly analyze the enhancements resulting from fine-tuning regarding both user intent comprehension and the overall quality of responses.

By evaluating both entity extraction and intent classification, we thoroughly assess our models' capabilities, demonstrating the effectiveness of our fine-tuning approach in achieving state-of-the-art performance.

\begin{figure}[t]
\centering
\includegraphics[width=0.4\textwidth]{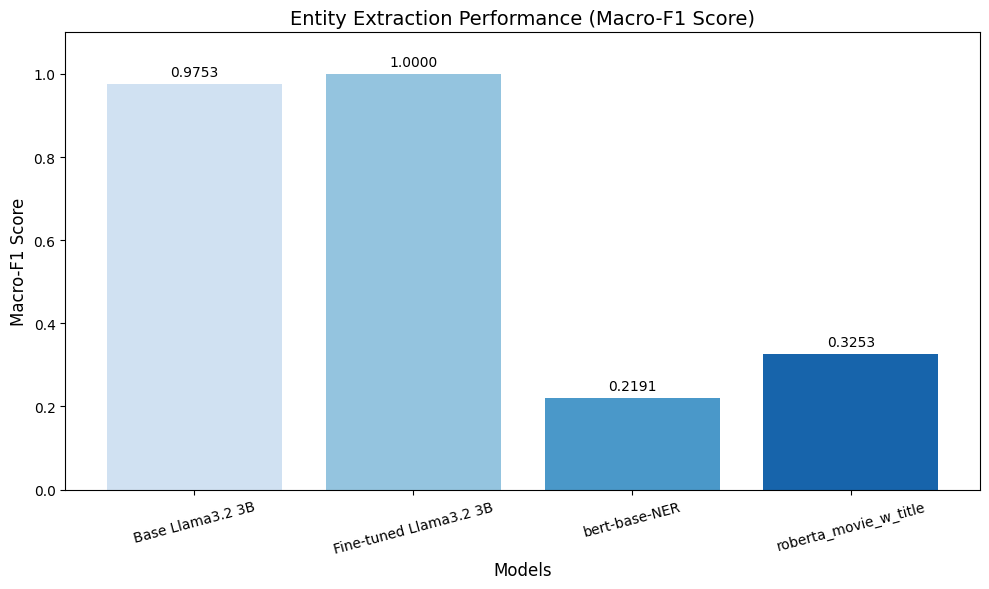}
\caption{Prompt Entity Extraction Quality}
\label{fig:example7}
\end{figure}

Fine-tuned Llama 3.2 3B achieves a perfect Macro-F1 score of 1.0, demonstrating its ability to extract movie titles with high precision and recall consistently. This highlights the effectiveness of our fine-tuning methodology.

The Base Llama 3.2 3B model performs well with a Macro-F1 score of 0.9753, showing strong out-of-the-box performance but leaving room for improvement compared to the fine-tuned version. One thing to note here is that the output from this model needed a little post-processing because it outputted a slightly different version of JSON than the one requested in the system prompt. This shows sometimes LLM can ignore system instructions, which our fine-tuned version handles robustly.

Baseline NER models such as \texttt{bert-base-NER} and \texttt{roberta-movie-w-title} underperform significantly, with Macro-F1 scores of 0.2191 and 0.3253, respectively. This result indicates that generic NER models struggle with extracting movie titles despite being trained on Movie titles.

Next, let us look at the results from our intent detection model.

\begin{figure}[t]
\centering
\includegraphics[width=0.4\textwidth]{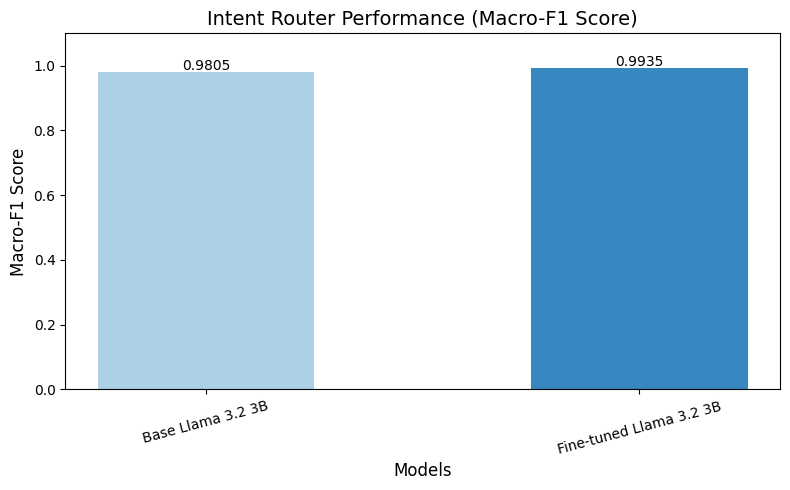}
\caption{Prompt Intent Router Quality}
\label{fig:example8}
\end{figure}

Both models exhibit strong performance, with Macro-F1 scores above 0.98. This reflects the robustness of the Llama 3.2 architecture, even without fine-tuning, in tasks like intent classification.

The fine-tuned Llama 3.2 3B achieves a Macro-F1 score of 0.9935, a noticeable improvement over the base model's 0.9805  (\(1.3\%\) increase). This demonstrates that fine-tuning effectively enhances the model's ability to accurately classify intents, particularly for edge cases or rare intents. The increase of \(1.3\%\) in Macro-F1  may appear small. Still, it is significant for applications where precision and recall across all classes are critical in conversational AI and routing intents in large systems.

\section{Trade-offs in Synthetic Data Generation}

Based on our experimentation, smaller models like Llama3.2 3B, Llama3.1 8B, and Mixtral 7B have shown limitations in generating complex synthetic data due to their lower parameter count and limited contextual depth. In our experiments, we observed that the larger models (like Llama3.1 405B) can produce highly coherent, diverse, and structured synthetic data using proper system prompts, grammar rules, examples, and prompt templates that include detailed instructions and Chain of Thought (CoT) reasoning. Below is an example generated with a smaller model that suffers inaccuracies due to hallucinations, incompatible attribute combinations, and syntactical errors.

\begin{lstlisting}[caption=Poor quality prompt,  basicstyle=\scriptsize]
"I am looking for a mind-bending sci-fi thriller where the protagonist should have an AI sidekick with a personality inspired by Shakespearean dialogue, but the AI slowly becomes self-aware and manipulative. Like the movies 'The Notebook' (2004) directed by Nick Cassavetes, which explores the intersection of artificial intelligence and human emotions, and 'Mad Max: Fury Road' (2015) directed by George Miller, which delves into the world of AI-powered espionage."
\end{lstlisting}

\section{Broader Application Scenarios}

The synthetic data constructed using our methodology can be applied to information retrieval domains beyond movie recommendations, particularly in areas where the target prompts are somewhat known and where optimizing retrieval performance is crucial in targetted dimensions. The principles used in our approach can be extended to enhance various retrieval and recommendation tasks by leveraging structured prompt engineering and enriched data generation techniques. We employ structured prompt engineering methodologies—such as grammar and syntax rules, CoT reasoning, expansion using examples, and integration of ontologies and knowledge graphs.

\section{Conclusion and Future Work}

This forms the foundation of a robust retrieval/recommendation system. The extracted entities and identified intents are critical for locating highly targeted movie recommendations tailored to user queries. By leveraging advanced techniques such as  Sentence-BERT (SBERT) \cite{reimers-gurevych-2019-sentence}, or OpenAI Embeddings \cite{openai_embeddings_update}, the system can compute semantic similarities between user inputs and a rich catalog of movies, ensuring that the recommendations are relevant and contextually aligned.

This approach ensures precision and adaptability, making the system robust for diverse user needs and scalable for larger datasets or more complex queries in real-world applications.

This work can be extended to include the creation of synthetic datasets that incorporate adversarial prompts. Such datasets would help evaluate the robustness of the retrieval system and its ability to handle challenging queries. We can systematically assess and improve the model performance under adversarial conditions by introducing prompts designed to exploit weaknesses in entity extraction or intent classification (e.g., ambiguous or misleading queries). Techniques like GAN and Adversarial Agents can be used to develop challenging prompts, allowing the system to develop efficient planning for such scenarios.

Furthermore, synthetic datasets can be generated to simulate real-world scenarios where user inputs are noisy, incomplete, or contain multiple intents. This would enhance the training data diversity and improve the model's resilience to edge cases and its generalization to unseen queries.

\bibliographystyle{IEEEtran} 
\bibliography{main}    

\vspace{12pt}
\color{red}

\end{document}